# Detecting a network of hijacked journals by its archive

Anna Abalkina

Ludwig Maximilian University of Munich

Anna.Abalkina@soziologie.uni-muenchen.de

## Abstract

The detection of hijacked journals is typically possible after some period of a clone website's operation. This study describes a method to detect hijacked journals based on the analysis of the archives of clone journals. This approach is most effective in discovering a network of hijacked journals that have the same organizer(s).

Analysis of the archives of clone journals allowed to detect 62 URLs of hijacked journals. It also provided the possibility to predict two clone websites before they became operational. This study shows that most detected hijacked journals represent a network of clone journals organized by one or several fraudulent individuals. The information and content of nine legitimate journals were compromised in international and national scientometric databases.

.

**Acknowledgments**. The author would like to acknowledge the assistance of Natalia Toganova and Evgeny Enikeev in writing the scripts.

**Conflict of interests**

The author has no conflicts of interest to declare that are relevant to the content of this article.

## Introduction and background

Clone journals became another manifestation of unscrupulous publishers. The first described case dates back to 2011 (Jalalian and Dadkhah 2015). Since then, hundreds of journals have been hijacked[1]. Hijacked journals mimic peer-reviewed journals (Lukić 2014; Bohannon 2015; Jalalian and Dadkhah 2015). They copy the ISSN and title of a legitimate journal to cheat potential clients (Dadkhah et al. 2016a) and provide a fake inflated impact factor (Jalalian and Mahboobi 2013; Jalalian 2015, Samuel and Aranha 2018). These journals target those researchers who are required to publish in journals indexed in international databases (Jalalian and Mahboobi 2014) such as Scopus or the Web of Science or those included in white lists, e.g., the UGC-CARE Approved list, which is applied in India.

Fraudulent publishers could use several methods to hijack a journal: register an expired domain (Jalalian and Dadkhah 2015, Memon 2019), hack the site of an authentic journal and register an alternative domain (Abalkina 2020a), or create a clone journal. Some fake clone journals cheat international databases by providing homepage links to fraudulent sites. Such cases were described by Jalalian and Dadkhah (2015) and Bohannon (2015) when some cyber links in Clarivate Master's journal list were directed to hijacked journals. Some hijackers were able to index the content of a clone journal in international databases (Abalkina 2020a; Al-Amr 2020).

The sophistication of cyber fraud and the frequency of successful attempts at fraud in international scientometric databases create a challenge for the international scientific community. Despite the fact that researchers, activists, and some institutions, such as the University Grant Commission in India, create lists of clone journals, there are no systematic checks of journal hijacking. In addition, systematic checks are hardly possible since the creation of a clone journal is an easy enterprise and the registration of the domain can be totally anonymous. Moreover, in most cases, it is possible to detect hijackers only after they use aggressive marketing and publishing. This creates a challenge to control hijackers and prevent the expansion of such a phenomenon in publishing.

[1] Hijacked journals are listed on the following websites: https://beallslist.net/hijacked-journals/, https://predatoryjournals.com/hijacked/, and https://ugccare.unipune.ac.in/Apps1/User/Web/CloneJournals

Hijacked journals and fraudulent publishers exploit the open access (OA) model and receive a fee for the publication of an article. Clone journals offer fast publication with no peer review. Clone journals menace academic integrity by publishing articles with no peer review, bias international database indexes, and become the repository of low-quality papers for a short period of time because clone websites of journals are not available after some period of time (Dadkhah et al. 2016b).

The interest in the study of hijacked journals was increased by the discovery of several clone journals in 2020. The fraudulent publishers hacked the site of the Talent Development and Excellence journal and cloned it. They also succeeded in indexing nearly 500 articles in Scopus in 2020 (Abalkina 2020a), which were later excluded after letters by the legitimate publishers. Unauthentic content of the Transylvanian Review also ended up in Scopus and later was deleted from the database (Al-Amr 2020). Another case was described on social media by the Russian researcher Eugene Osin. He found his paper cited in a clone of a German journal Waffen- und Kostümkunde. This discovery was possible because the article written by Russian scholars in the clone journal that cited Osin’s paper was indexed in the Russian scientometric database eLibrary.

The inspection of the low-quality clone website of a German journal raised several questions. First, for how long was this website working? Second, why did the archive section of the journal contained issues since 2014, despite their periodicity being different (biannual vs annual issues)? The check of the domain using the Whois service showed that the website was updated on January 8, 2020 (the registration date is not available). This approach was also used by Bohannon (2015) who detected hijacked journals using the recent registration of the sites and was advised by Asadi et al. (2016) to distinguish legitimate publishers and clone journals.

If the clone journal presumably started its activity in 2020, what about the archive content of the previous five to six years? The hypothesis was that content should be as fake as the journal itself. The expectation was to find articles that had already been published in other journals. The alternative hypothesis was that those naïve authors who were cheated published their original articles. Ten authors whose articles were found in the archive of the clone journal Waffen- und Kostumkunde were contacted. Only two answered, and they confirmed that they have never submitted a paper to the journal. Papers of these scholars were duplicated and published in the archive of the hijacked journal. This proved that the hypothesis of fake content was correct. A manual Google search of the title of the articles and authors allowed to detect several other clone journals that reused the same papers for their archive sections.

Dadkhah et al. (2016b) demonstrated evidence of the circulation of the same texts between hijacked and predatory journals. The articles in hijacked journals are not indexed, and text similarities cannot be detected by anti-plagiarism software (Dadkhah et al. 2016b). Hijacked journals accept articles without peer review, and it is very likely that they do not check texts for the presence of plagiarism/self-plagiarism. The migration of text similarities was also demonstrated by Russian Dissernet, which detected the network of dissertation mills and circulation of identical texts without a proper citation between numerous PhD theses (Abalkina 2020b). PhD theses defended in dissertation mills were reused and thus became a predictor of other dishonest dissertations. Based on this evidence, the content of hijacked journals can be a predictor of other fraudulent journals, especially if they belong to the same network.

In such fraudulent mills, hijacked journal text can only show the visibility of a scientific article and must comply with formal criteria, such as having a title, abstract, content and bibliography. I will study whether the archives of hijacked journals can be used to predict other hijacked journals, especially from the same network. To conduct the analysis, I need to search the titles and authors of the detected hijacked journals to find other hijacked journals from the network.

The remainder of this paper is organized as follows. The next section overviews the literature on the economic theory of crime, the circulation of the text similarities and the determinants to publish in low-quality journals. Section three presents the argument. Section four discusses the methods. Section five reports the results of the study. Section six provides the discussion. Section seven concludes.

## Literature review

In my study, I use insights and evidence from several literature streams. First, I examine the literature on the economic theory of crime and rational behavior of criminals, which depends on the costs and benefits of fraud. Second, I examine the determinants of publishing in low-quality journals. Third, I review the research on academic misconduct and plagiarism with a focus on the recycling of texts in predatory or hijacked journals. In this study, the terms ‘text recycling’ and ‘recycled content’ will be used in their broad senses to refer to both plagiarism and self-plagiarism.

According to the economic theory of crime criminals act rationally, their criminal behavior depends on the benefits of fraudulent behavior and the costs of the actions together with the probability of being caught and the severity of punishment (Becker 1968, Garoupa 2014). The benefits of fraudulent actions include income from the fraudulent act. Although the costs of fraudulent actions include the costs of organizing the fraudulent action and the costs of punishment, moral values and psychological factors are also considered. According to the economic theory of crime, fraud will take place if the benefits exceed the costs. One of the consequences of this economic theory of crime suggests that criminals would minimize their costs.

Hijacked journals represent a type of cybercrime. Cybercriminals cheat the scholars who wish to publish articles in journals included in different white lists. As the average time of such fraud is not long due to possible detection of a clone journal, hijackers will try to reduce the costs and maximize the benefits.

The success of fraud also depends on the demand of scholars for publications. Scholars are required to publish papers in order to meet the requirements to accomplish PhD theses, career promotions or grant applications (De Bond and Miller 2005; Dyke 2019; Huang 2020). Publication pressure increases the probability of cheating by scholars due to the increased benefits of fraudulent behavior. Publish or perish pressure has been found to be a factor that increases the probability of academic misconduct by scholars (Martinson et al. 2005; Necker 2014; Fanelli et al.2015; Hopp and Hoover 2017; Burton et al. 2020). The evidence also shows that poor performers would increase cheating under competitive pressure (Schwieren and Weichselbaumer 2010; Cartwright and Menezes 2014).

Hijacked journals exploit this publish or perish culture in order to attract and cheat scholars who wish to publish articles in order to meet the requirements for publications. There are four main groups of factors why scholars choose to publish in low-quality journals that offer the fast publication of a paper. The first group of factors is connected with the academic requirements for publications, e.g., to apply for a job or to obtain academic promotion (Seethapathy et al. 2016; Demir 2018; Cortegiani et al. 2020), to pass regular academic attestation (Bagues et al. 2019), to fulfil the requirements of publications for a PhD degree (Seethapathy et al. 2016; Patwardhan 2019), and to obtain a financial bonus for a publication (Demir 2018). The second group of factors explains the publication in low-quality journals by young or unexperienced researchers who may be unaware of the quality of the journals and submit papers to them (Kurt 2018). The third group is related to the quality of the research itself. There is also evidence that some researchers are unable to provide high-quality research that is required for prestigious journals (Demir 2018). Poor proficiency in English is also a reason that researchers do not submit papers to high-quality journals (Kurt 2018). The last group is connected with social or religious identity. Researchers belonging to developing counties or to certain religions do not believe that their papers would be treated equally or even considered when submitted to prestigious journals (Kurt 2018).

Predatory and hijacked journals are prone to academic misconduct and dishonesty. They accept the publication of papers containing serious errors (Bohannon 2013), pseudoscientific ideas (RAS 2020) or plagiarism. Owens and Nicoll (2019) examined the content of three predatory nursing journals in order to detect plagiarism or duplication of papers. They studied 296 articles, and in 100 of them (68%), they found exact or near exact text similarities. Abad-García (2019) also provides evidence from fraudulent scholars who copied the articles of other authors and published them under their own names in predatory journals. I have already mentioned that there is evidence of similar text circulation between predatory and hijacked journals (Dadkhah et al. 2016b). These observations provide further support for the hypothesis that low-quality journals are predisposed to plagiarism and academic dishonesty.

There is currently considerable concern about predatory and hijacked journals that provide poor quality or no peer review and do not check enough or any of the texts for plagiarism. This creates a challenge for academic integrity. Furthermore, we still do not know enough about plagiarism in journals because the knowledge on text similarities depends directly on systematic checks for plagiarism, which are not done yet by the scientific community due to obvious time- and resource-consuming procedures and challenges in detecting text similarities (Weber-Wulff 2019). This creates opportunities for poor-quality journals and hijacked journals to develop their business with impunity and without regard to academic integrity.

## Argument

The business of hijacked journals consists of mimicking an original journal, including copying the name, the ISSN, and the journal's reputation, as well as its indexes in international databases. Hijacked journals collect fees for fast publication without peer review. The amount of the fee depends on the reputation of the original journal and on the quality of the hack. According to the evidence of victims, the fees could reach up to $1000 for a paper in a clone journal that was able to index the article in the Scopus database.

According to the economic theory of crime, fraudulent publishers maximize the utility function given the probability that the fraud is detected and punished. Journal hijacking cybercrimes have a very low risk of punishment. The probability of detection of the owners of such fraudulent businesses is not high. Most of the domains that host hijacked journals are anonymous according to the Whois service data.

Using the economic theory of crime approach, fraudulent publishers should minimize the costs of detecting the fake status of the journal. Fraudulent publishers tend to mimic all formal criteria by copying the ISSN and title of a journal, which are essential parts of the business. Some hijackers change some words in the title of the journal to avoid legal accusations of the theft of the journal title. In many cases, the words “the Journal of” or “Research journal of” are added (Abalkina 2020a).

One of the important features of a reputable journal is its archive. A newly created clone journal has four alternatives in this case: (a) to leave empty content, (b) to use the content of an authentic journal, (c) to create original content of the journal, and (d) to create recycled content. Hijacked journals normally do not copy the content of authentic journals (with some exceptions). This is connected with the fact that the latter publishes papers on specific subareas of research and/or in native languages. Many hijacked journals position themselves as multidisciplinary and publish articles mostly in English. That is why it is not promising to use the original content. There are also legal considerations to copying papers from authentic journals. It is not possible to do without the permission of a legitimate journal (Sanderson 2010). Furthermore, the content of a reputable journal can be tracked, and the hijacked journal can be easily detected.

The content of the journal and the continuity of publishing are the factors that can attract potential clients by creating an illusion of an authentic journal with archive issues. In this case, leaving the content blank will not attract potential clients because they most likely will not consider an empty journal as an authentic journal (Dony 2020).

In this case, two alternatives are left for fraudulent publishers: to create original content or to use recycled texts. Fraudulent journals will not create original content due to its high costs and will use recycled text with higher probability. Clone journals would use the less expensive option to create the content. Fraudulent publishers can take the text from hijacked or predatory publishers but not legitimate journals because the first two are not well known to readers, and it is more difficult for authentic authors or readers to detect fraud. The recent finding by Björk et al. (2020) showed that 56% of the articles in predatory journals receive zero citations while only 9% of the articles indexed in international databases are not cited, e.g., papers from predatory journals are less visible to the audience. The content from hijacked journals also has limited visibility. Moreover, due to the short operating period of hijackers, the content can be lost; thus, the articles published in hijacked journals can be lost for the scientific community (Dadkhah and Borchardt 2016; Van Zundert and Klein 2019), and the recycling of such content cannot be detected after some period of time. In this case, the archive of a newly created clone journal is replenished by texts from hijacked or predatory journals created via text recycling.

The evidence from Russian plagiarised PhD theses shows that within a dissertation mill each text of the thesis can be reused plenty of times, and thus other dishonest PhD holders can be detected (Abalkina 2020b). By searching the content of the journal (titles and authors or titles) it could be possible to detect the journals where such content was published and re-cycled. If our hypothesis is correct we can define the whole network of hijacked journals using the method of searching the fake archive of the hijacked journals. Such snowball method can be mostly efficient when the network organization of a fraudulent business is detected, e.g. the content of new PhD theses, newly organized hijacked journals is systematically based on plagiarism from the texts already present within the network.

## Methodology

The main goal of this study is to investigate the whole network of hijacked journals by the same organizer(s). I hypothesize that hijacked journals, especially from the same network, share identical texts in archives. The goal of the research is to detect hijacked journals on the basis of the same texts. I hypothesize that fraudulent publishers will not change the titles of the articles in order to create a fictitious archive of the journal. For this case, the search of the titles of articles or the titles of articles with authors will show the link where such a text was published.

On the basis of the argument, the methodology was developed. The initial sample of the hijacked journals consisted of several items that were detected manually by searching the content of the clone of Waffen- und Kostümkunde. I extracted articles' information like title or title together with authors and authors' affiliation of articles depending on how this data is presented on the website of a hijacked journal (see Appendix 1). This information found in the archive of the clone journals was extracted using a script. Another script searched the extracted data using a *Google Custom Search API*, and the first ten Internet addresses were memorized. I manually tested how many exact titles were shown in the Google search. The maximum result was six, and I added four more to account for a possible bias.

Then, all Internet links of the search were arranged in alphabetic order. The websites that appeared many times were checked to determine if the journal sites were clones of legitimate journals. Using the snowball method, the archive of the newly detected hijacked journals was analyzed. When the number of detected hijacked journals' URLs reached 58 I joined all search results and again arranged them in alphabetical order. It allowed to detect several more clone journals. The search was stopped when no new websites of hijacked journals were found. The search was conducted in October 2020 – January 2021. The titles only in the English language were searched. The archive of 61 hijacked journals' URLs was checked, and 57051 research requests were made.

## Results

It was hypothesized that clone journals also use cloned texts. Fraudulent publishers reduce costs and use the same articles to form the fake archives of clone journals. The current study of the content of hijacked journals detected 62 URLs of 57 hijacked journals. Most of the detected clone journals that represented the same network of hijacked journals had identical articles in their archives. At least one stand-alone hijacked journal was identified due to the publication of plagiarized articles or 'predatory' authors, and one journal was found due to the identical boards of editors of both a legitimate and another hijacked journal. In most cases, the hijackers used the same texts and just changed the title of the corresponding journal, but in some cases, they failed to even change the titles. The list of hijacked journals obtained from the analysis is presented in Appendix 2.

I didn't include in the research results a number of cases. In the first one it was not possible to distinguish between the legitimate publisher of "International Journal of Advanced Science and Technology" and some other journals of the group and their clones. The second case concerns the journal "Indian Journal of Forensic Medicine & Toxicology". The alternative website of the journal (medicopublication.com) contains the identical archive of a legitimate journal. The third case is the website tifanjournal.com. With the high probability this website belonged to the network of the hijacked journals but as of December 2020 the domain has expired. There is also the domain pjrpublication.com, it belongs to the network of detected hijacked journals but all links inside this website are redirected to the clone of the Paideuma Journal of Research (paideumajournal.com). Another hijacked journal most likely belonging to the network is Dogo Rangsang Research Journal (www.dogorangsang.org) but the website is not filled yet and doesn't contain any issue. This website was discovered during the spellchecking of the original journal which had been hijacked earlier by an alternative website. I was also not able to detect the legitimate publisher of the TAGA Journal of Graphic Technology (www.tagajournal.com). According to ISSN Portal the website www.tagajournal.com is the official URL of TAGA Journal but the titles of the journals don't coincide. The TAGA Journal of Graphic Technology has a fake editorial board, it coincides with editorial team of Electronic Journal of Graph Theory and Applications. The website of the TAGA Journal of Graphic Technology has the same registrant contact as several hijacked journals from the network.

The most obvious finding to emerge from the analysis is that hijacked journals represent the network of clone journals that are registered regularly by cybercriminal(s). Most of these hijacked journals are not stand-alone journals. This may indicate the emergence of systematic journal hijacking. The graph in Figure 1 represents the links between journals that share mostly identical texts. Each edge of the graph is created if a website of a clone journal is found in the search results of titles/authors/authors' affiliation of another clone journal using a *Google Custom Search API.* The graph has a single connected component and it is expectable because a snowball method has been applied. This graph has the average degree of 33,8, it means that on average each clone journal has links with other 33 clone journals. It seems that the same texts were shuffled like a deck of cards and distributed among the clone journals.

This business model of the network organization can also be observed in PhD theses mills and paper mills. For example, in Russia, PhD theses mills are based on the circulation of the same texts, which are defended many times by different 'scholars,' and the involvement of the same professors who organized dishonest defenses (Abalkina 2020b). The same texts can be used by different fraudulent factories of plagiarized PhD theses.

The first case of this network of hijacked journals was recorded in 2017 when the domain of the hijacked Journal for Advanced Research in Applied Sciences was registered. Since 2020, the frequency of hijacked publishing has grown. Approximately 50% of hijacked journal domains were registered (or in several cases, updated) in 2020 (see Figure 2). Nine clone journals were registered on 8 January 2020 alone.

**Figure 1**

## The network of clone journals

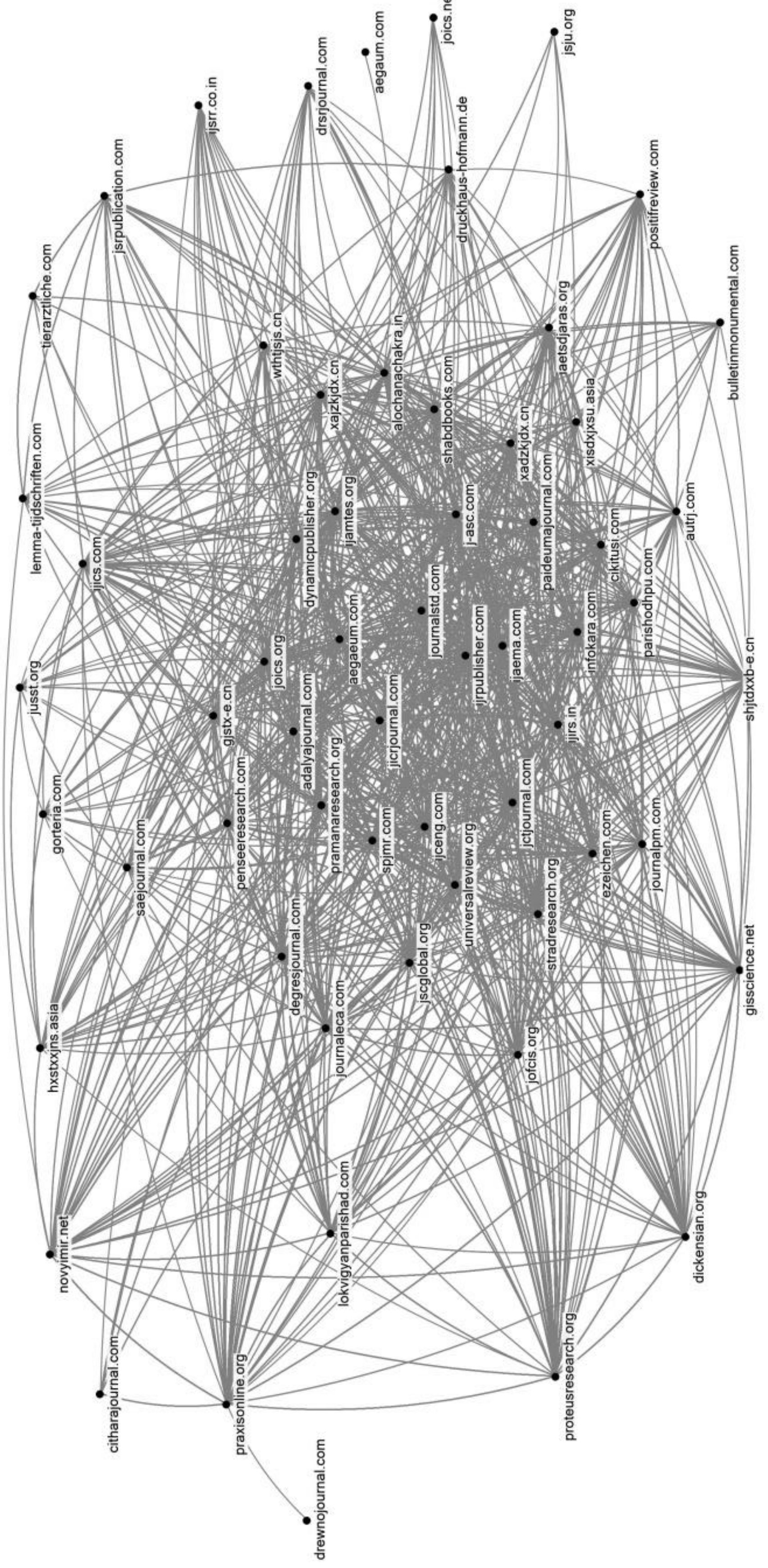

There is also an interesting case of the Cithara journal (citharajournal.com). According to the Whois service, the domain was registered in 1995. It seems that cybercriminals registered the expired domain and updated it on 7 January 2020. The clone journal itself became operative approximately at the beginning of November 2020. Its website always appeared in the research results earlier, but the main webpage did not look like the website of a journal. There was also one nonoperative website of the clone "Drewno Journal" (the domain was registered on 18 August 2020). It already had fake archive content that was detected during the analysis, and it seems that the clone journal will start to cheat authors soon. As of 31 December 2020 the hijacked "Drewno Journal" is not operative.

**Figure 2**

**Date of registration of the domain of clone journals**

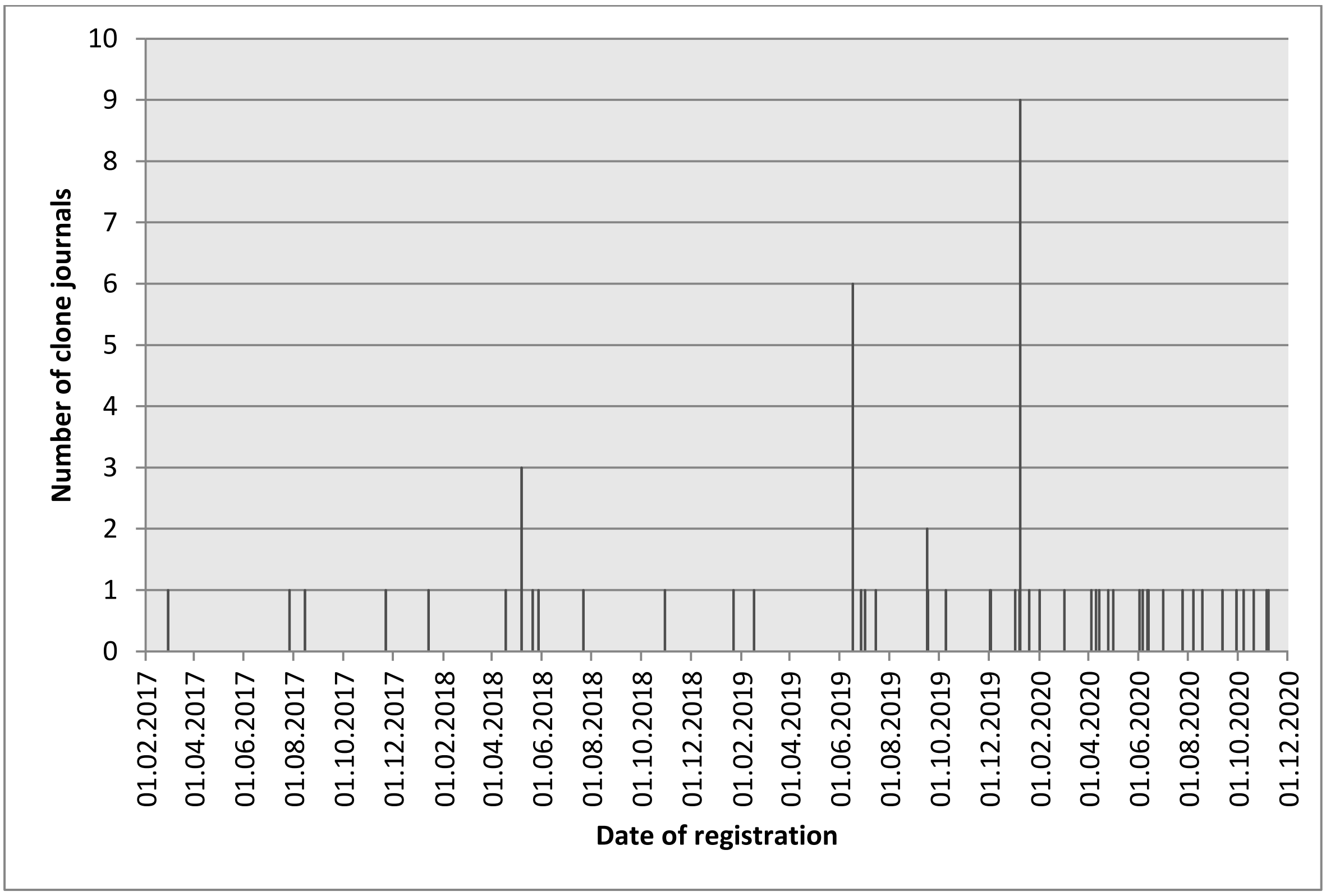

* - For the purpose of the figure, the date of registration of the Cithara journal is considered to be 7 January 2020.

Most of the detected journals belong to one network and are created by the same fraudulent individual(s) or group(s) of individuals. The hijacked journals of the network are created according to the same scheme. Most clone journals (with several exceptions) use the identical ISSN number of the original journals on their websites. The titles of the clone journals are slightly changed. Some letters are changed or some additional words such as “Journal”, “Journal of”, “Multidisciplinary Journal”, “Research” or “Research journal of” are added. The clone websites have similarly structured webpages, and some websites are registered on the same day.

The boards of editors of the journals are fake and have been copied from international legitimate journals. Such reputable members of the board should attract potential clients, but surprisingly, it can also increase the visibility of clone journals. The search of the provenance of the board of editors of the clone journals allowed to detect another clone journal, Aegaum Journal (aegaum.com). The probability of detecting this journal using content analysis was low due to the archive containing just 12 articles (as of December 2020 – 11 articles).

All hijacked journals demonstrated fake and inflated impact factors in the range from 1.07 to 6.3 (the most frequent impact factor was 6.1). The contact emails indicated on the clone websites are mostly from free email services. The DOIs of the articles are also fake, not only because they are not registered but also because some of them do not start with ‘10’, as all DOIs should.

It is difficult to detect original papers and recycled papers published in clone journals. However, the authors of the papers published in hijacked journals mainly represent India and other developing countries.

Previous research has noted that hijackers create clones of journals published in non-English languages or by small publishers (Jalalian and Mahboobi 2014, Shahri et al. 2016). The analysis of the authentic journals that were hijacked by detected clone journals confirmed the previous findings but also allowed to define more specific groups of journals.

- Niche journals on art, archeology, weapons, literature, etc. This strategy to hijack niche journals decreases the probability of being quickly detected. Though clone journals are multidisciplinary, their content does not correspond to the subject area of the original journal.

The authentic journals are published in English or a native language. During 2020, a number of Chinese journals were hijacked.

The target journals can even be a national cultural heritage in their country of origin, such as Novyi Mir in the USSR/Russia. It is a literature journal that first published “One Day in the Life of Ivan Denisovich” by Nobel laureate Aleksandr Solzhenitsyn. This journal is indexed in the Web of Science, and now it is mimicked by fraudulent publishers as a multidisciplinary journal.

- Another category of authentic journals are print journals that do not have URLs. These authentic journals that have been hijacked in the network are mainly published in India.

- Journals that stopped publishing. Hijackers cloned several journals that stopped publishing or changed their titles. For example, the authentic Journal of Interdisciplinary Cycle Research was renamed Biological Rhythm Research in 1994. Fraudulent criminal cloning used the old title and ISSN of the journal to cheat potential clients. The journal ‘Science, Technology and Development’ ceased activity, but its data were cloned. This strategy allows cybercriminals to avoid the legal consequences of using the title of an authentic journal.

- Journals that are/were indexed in Scopus or the Web of Science. The clones of these journals increase the probability of attracting potential clients who are required to publish in indexed journals.

Another interesting finding of the study was that the research results showed many links to predatory journals. This supports the previous finding by Dadkhah et al. (2016b) that texts published in hijacked journals could be plagiarized from other researchers. This study extends the knowledge of the patterns. First, this research confirms the presence of plagiarized text in hijacked journals originating from predatory publishers. Second, hijacked journals published identical texts (with the correct authorship) taken from predatory journals to replenish their archives. Third, I found that the same authors had articles both in predatory and hijacked journals. This could suggest the pattern of ‘predatory’ authors in these journals.

The most striking aspect of the detected hijacked journals is in the links to the clone websites and the articles from clone journals penetrating different legitimate scientometric databases (see Table 1).

**Table 1**

**Clone journals in scientometric databases**

| | **Scopus** | **Scimago** | **eLibrary** |
|---|---|---|---|
| Link to the website of the clone journal | **1** | **3** | **2** |
| Indexed content from the clone journal | **1** | **-** | **5** |
| Total number of journals with a link to the clone website and/or indexed content from the clone journal | **1** | **3** | **5** |

I found at least three cases of links to clone sites in Scimago. I also detected one case in Scopus where the link to homepage was directed to a clone journal, and the same journal had indexed content in Scopus. The same problem was identified with the Russian database eLibrary, where two journals had incorrect homepages and five journals had fake content in the database. Overall, the information of nine journals (~16%) in different scientometric databases was compromised.

These findings may help to understand why hijacked journals are so successful in cheating authors. Hijacked journals create a serious menace for academic integrity. As commented on in the Scimago webpage of the hijacked “Talent Development and Excellence” journal, some content ended up in Scopus where “researchers were not investigators” and were unable to verify the information from a legitimate scientometric database[2]. It is sometimes not easy to distinguish authentic and clone journals, even for professionals.

## Limitations and discussion

An initial objective of the project was to identify the network of hijacked journals by analyzing the archive of the detected clone journals. This detection method is most effective in identifying hijacked journals that belong to the same network of clone journals. Due to the circulation of texts between hijacked journals, this method also allowed the identification at least one stand-alone hijacked journal: the Journal of Southwest Jiaotong University (jsju.org). This journal and the hijacked Journal of Composition Theory published nearly identical text but signed by different authors.

In addition to hijacked journals, predatory journals were among the search results. Similar results were detected between articles published in hijacked journals and predatory journals. There have also been authors who had their papers published in both predatory and hijacked journals. 'Predatory' authors could be a good predictor of poor-quality journals. The study by the RAS Commission for Counteracting the Falsification of Scientific Research showed that Russian scholars who are mostly involved in publications in predatory journals and plagiarism cases predict predatory journals well (RAS 2020). These findings raise questions regarding the determinants of publications in hijacked journals.

Currently, there is still no evidence on the reasons authors publish in hijacked journals, although it is believed that naïve or unexperienced scholars are victims of cybercrime fraud. However, is this the only category of scholars who publish in hijacked journals? We do not have much supportive evidence that the authors are aware of hijacked journals. This could be an important issue for future research on the market of hijacked or predatory journals due to the significant rise in poor-quality or fraudulent journals. There is evidence of high demand for publications in predatory journals to meet the requirements of universities or to obtain financial bonuses. One of the future research topics can be to test the hypothesis that authors can recognize fraudulent journals but still publish in them to cheat the system, to inflate their publication records, etc. because ‘predatory’ authors can be attracted by fraudulent publishers. Another possible research topic is the prediction of hijacked and predatory journals by ‘predatory’ authors.

[2] https://www.scimagojr.com/journalsearch.php?q=21100208309&tip=sid&clean=0

## Conclusions

Despite the limitations, this study contributes in the following ways. First, this study explores a new method of identifying hijacked journals. Most of these journals belong to the same network, e.g., they have the same organizer(s). Hijacked journals from the same network recycle texts to demonstrate continuous publishing in order to cheat potential clients. To date, hijacked journals can be tracked by the Whois service (Bohannon 2015) or can be distinguished by classification algorithms (Dadkhah et al. 2016c; Shahri et al. 2018). This new approach allowed to detect 62 URLs of 57 hijacked journals. Most of these URLs have not been included in the available lists of hijacked or clone journals. This method also allowed to detect at least one stand-alone clone journal. Second, this study has allowed to predict two clone journals before their websites became operational. Before this study, it seems that all hijacked journals were detected *post factum*. Third, this research improves the current knowledge on the circulation of text similarities. The same texts are circulated between hijacked journals and between hijacked and predatory journals. Fourth, this study confirms the evidence on the target journals that are cloned and expands the evidence on them. Fifth, this study shows that the data and content of nine journals have been compromised in legitimate scientometric databases.

## References:

1. Abad-García, (2019). Plagiarism and predatory journals: A threat to scientific integrity. Anales de Pediatría, Volume 90, Issue 1, Pages 57.e1-57.e8. https://doi.org/10.1016/j.anpede.2018.11.006

2. Abalkina A. (2020a). The case of the stolen journal. Retraction Watch, July 7. URL: https://retractionwatch.com/2020/07/07/the-case-of-the-stolen-journal/ (retrieved on 20.10.2020).
3. Abalkina A. (2020b). Organisation of dissertation mills in Russia. 6th International Conference PAEB 2020 First Virtual ENAI Conference.
4. Al-Amr M. (2020). How did content from a hijacked journal end up in one of the world's most-used databases? Retraction Watch. September 1. URL: https://retractionwatch.com/2020/09/01/how-did-content-from-a-hijacked-journal-end-up-in-one-of-the-worlds-most-used-databases/ (retrieved on 20.10.2020).
5. Asadi, A., Rahbar, N., Asadi, M. et al. (2017). Online-Based Approaches to Identify Real Journals and Publishers from Hijacked Ones. Science and Engineering Ethics 23, 305–308. https://doi.org/10.1007/s11948-015-9747-9
6. Bagues M., Sylos-Labini M., Zinovyeva N. (2019). A walk on the wild side: 'Predatory' journals and information asymmetries in scientific evaluations. Research Policy, Volume 48, Issue 2.
7. Becker G.(1968). Crime and Punishment: An Economic Approach. Journal of Political Economy. Vol. 76, No. 2 (Mar. - Apr., 1968), pp. 169-217
8. Björk, B.-C.; Kanto-Karvonen, S.; Harviainen, J.T. How Frequently Are Articles in Predatory Open Access Journals Cited. Publications 2020, *8*, 17.
9. Bohannon, J. (2015). How to hijack a journal. Science, 350(6263), 903–905.
10. Bohannon, J. (2013). Who's afraid of peer review? Science, 342 (6154), 60-65.
11. Bruton S. V., Medlin M., Brown M., Sacco D. F. (2020). Personal motivations and systemic incentives: Scientists on questionable research practices. Science and Engineering Ethics, 26, 1531–1547.
12. Cartwright E., Menezes M-. (2014). Cheating to win: Dishonesty and the intensity of competition, Economics Letters, Volume 122, Issue 1, Pages 55-58.

13. Cortegiani A., Manca A., Giarratano A. (2020). Predatory journals and conferences: why fake counts. Current Opinion in Anaesthesiology. Vol. 33, Issue 2, 192-197.
14. Dadkhah M, Borchardt G. (2016). Hijacked journals: an emerging challenge for scholarly publishing. *Aestheth Surg J* 2016; 36:739–741.
15. Dadkhah, M., Maliszewski, T., Jasi, M. (2016a). Characteristics of Hijacked Journals and Predatory Publishers: Our Observations in the Academic World. Trends in Pharmacological Sciences. Volume 37, Issue 6, June 2016, Pages 415-418.
16. Dadkhah, M., Maliszewski, T. & Teixeira da Silva, J.A. (2016b). Hijacked journals, hijacked web-sites, journal phishing, misleading metrics, and predatory publishing: actual and potential threats to academic integrity and publishing ethics. Forensic Sci Med Pathol 12, 353–362. https://doi.org/10.1007/s12024-016-9785-x
17. Dadkhah, M., Maliszewski, T., & Lyashenko, V. V. (2016c). An approach for preventing the indexing of hijacked journal articles in scientific databases. Behaviour & Information Technology, 35(4), 298–303.
18. De Rond M., Miller A. N. (2005). Publish or Perish. Journal of Management Inquiry, 14 (4), 321–329. doi:10.1177/1056492605276850
19. Demir S.B. (2018). Predatory journals: Who publishes in them and why? Journal of Informetrics. Volume 12, Issue 4.
20. Dony C., Raskinet M., Renaville F., Simon S., Thirion P. (2020). How Reliable and Useful Is Cabell’s Blacklist? A Data-Driven Analysis. Liber Quarterly, Vol. 30, 1–38
21. Dyke G. (2019). Does the early career ‘publish or perish’ myth represent an opportunity for the publishing industry? Learned Publishing, 32, 90–94.
22. Fanelli, D., Costas, R., Lariviere, V. (2015). Misconduct policies, academic culture and career stage, not gender or pressures to publish, affect scientific integrity. PLOS ONE 10 (6), e0127556.

23. Garoupa N. (2014) Economic Theory of Criminal Behavior. In: Bruinsma G., Weisburd D. (eds) Encyclopedia of Criminology and Criminal Justice. Springer, New York, NY.
24. Huang Y.(2020). Doctoral writing for publication, Higher Education Research & Development. https://doi.org/10.1080/07294360.2020.1789073
25. Hopp C., Hoover G.A. (2017). How prevalent is academic misconduct in management research? Journal of Business Research, 80, 73-81.
26. Jalalian, M. (2015). The Story of Fake Impact Factor Companies and How We Detected Them. Electronic Physician 7 (2): 1069–1072.
27. Jalalian M., Dadkhah M. (2015). The full story of 90 hijacked journals from August 2011 to June 2015. Geographica Pannonica, 19(2), 73–87.
28. Jalalian M., Mahboobi J. (2013). New corruption detected: bogus impact factors compiled by fake organizations. Electronic Physician; 5: 685– 686. doi: 10.14661/2013.685-686
29. Jalalian M., Mahboobi J. (2014). Hijacked Journals and Predatory Publishers: Is There a Need to Re-Think How to Assess the Quality of Academic Research? Walailak Journal of Science and Technolgy, 11(5).
30. Honig B., Bedi A. The Fox in the Hen House: A Critical Examination of Plagiarism Among Members of the Academy of Management. Academy of Management Learning & Education, 2012. Vol. 11, No. 1, P. 101–123.
31. Martinson B., Anderson M., de Vries R. (2005). Scientists behaving badly. Nature 435, 737–738. https://doi.org/10.1038/435737a
32. Memon A. (2019). Hijacked journals: A challenge unaddressed to the developing world. Journal of the Pakistan Medical Association. 69(10):1413-1415.
33. Necker S. (2014). Scientific misbehavior in economics. Research Policy. Volume 43, Issue 10, Pages 1747-1759.

34. Owens J.K., Nicoll L.H. (2019). Plagiarism in Predatory Publications: A Comparative Study of Three Nursing Journals. Journal of Nursing Scholarship, 2019; 51:3, 356–363.
35. Patwardhan B. (2019). Why India is striking back against predatory journals. Nature 571, 7.
36. RAS (2020). Inostrannyye khishchnyye zhurnaly v Scopus i WoS: perevodnoy plagiat i rossiyskiye nedobrosovestnyye avtory [Predatory Journals at Scopus and WoS: Translation Plagiarism from Russian Sources. Commission for Counteracting the Falsification of Scientific Research] in collaboration with Anna A. Abalkina, Alexei S. Kassian, Larisa G. Melikhova.
37. Samuel A.J., Aranha V.P. (2018). Valuable research in fake journals and self-boasting with fake metrics. Journal of Pediatric Neurosciences 13 (4).
38. Sanderson K. (2010). Two new journals copy the old. Nature 463, 148.
39. Schwieren, C., Weichselbaumer, D. (2010). Does competition enhance performance or cheating? A laboratory experiment. Journal of Economic Psychology 31 (3),241–253. Nature 463, 148. doi:10.1038/463148a
40. Seethapathy, G., Kumar, J., & Hareesha, A. (2016). India's scientific publication in predatory journals: Need for regulating quality of Indian science and education. Current Science, 111(11), 1759-1764.
41. Shahri et al. (2018). Detecting Hijacked Journals by Using Classification Algorithms. Science and Engineering Ethics, 1–14.
42. Van Zundert A., Klein A. (2019). How to avoid predatory and hijacking publishers? European Journal of Anaesthesiology, Vol.36, Issue 11, 807-809.
43. Weber-Wulff D. (2019). Plagiarism detectors are a crutch, and a problem. Nature, 567.

**Appendix 1**

## Number of search queries and type of search of clone journals

| N | Journal (clone) | URL (clone) | ISSN (clone) | Number of search queries | Type of search |
|---|---|---|---|---|---|
| 1 | Adalya Journal | http://adalyajournal.com/ | 1301-2746 | 1747 | Title of articles/authors/ affiliation |
| 2 | Aegaeum Journal | http://aegaeum.com/ | 0776-3808 | 1305 | Title of articles/authors/ affiliation |
| 3 | Aegaum Journal | https://aegaum.com/ | 0776-3808 | 12 | Title of articles/authors |
| 4 | Alochana Chakra Journal | http://www.alochanachakra.in/* | 2231-3990 | 2183 | Title of articles/authors/ affiliation |
| 5 | Aut Aut Research Journal | http://autrj.com/ | 0005-0601 | 361 | Title of articles/authors/ affiliation |
| 6 | Bulletin Monumental Journal | http://bulletinmonumental.com/ | 0007-473X | 147 | Title of articles/authors/ affiliation |
| 7 | Cikitusi Journal for Multidisciplinary Research | http://www.cikitusi.com/ | 0975-6876 | 644 | Title of articles/authors/ affiliation |
| 8 | Cithara Journal | http://citharajournal.com/ | 0009-7527 | 67 | Title of articles/authors/ affiliation |
| 9 | Compliance Engineering Journal | http://ijceng.com/ | 0898-3577 | 1153 | Title of articles/authors/ affiliation |
| 10 | Degres Journal | https://degresjournal.com/ | 0376-8163 | 324 | Title of articles/authors |
| 11 | Dogo Rangsang Research Journal (D.R S. Research Journal) | http://www.drsrjournal.com/ | 2347-7180 | 641 | Title of articles |
| 12 | Drewno Journal | https://drewnojournal.com/** | 1644-3985 | - | - |
| 13 | G & O (Gedrag & Organisatie) | http://lemma-tijdschriften.com/ | 0921-5077 | 282 | Title of articles/authors |
| 14 | GIS Science Journal | http://www.gisscience.net/ | 1869-9391 | 313 | Title of articles/authors/ affiliation |
| 15 | Gorteria Journal | https://gorteria.com/ | 0017-2294 | 259 | Title of articles |
| 16 | High Technology Letters | http://www.gjstx-e.cn/ | 1006-6748 | 727 | Title of articles/authors |
| 17 | Infokara Research | http://www.infokara.com/ | 1021-9056 | 1242 | Title of articles/authors/ affiliation |
| 18 | International Journal of Information and Computing Science | http://ijics.com/ | 0972-1347 | 1491 | Title of articles/authors/ affiliation |
| 19 | International Journal of Innovative Research & Studies | http://ijirs.in/ | 2319-9725 | 1226 | Title of articles/authors/ affiliation |
| 20 | International Journal of Management, Technology and Engineering | http://www.ijamtes.org/ | 2249-7455 | 2277 | Title of articles/authors/ affiliation |
| 21 | International Journal of Research | http://ijrpublisher.com/ | 2236-6124 | 2975 | Title of articles/authors/ affiliation |
| 22 | International Journal of Scientific Research and Review | http://www.dynamicpublisher.org/ | 2279-543X | 2164 | Title of articles/authors/ affiliation |
| 23 | International Journal of Scientific Research and Review | http://www.ijsrr.co.in/ | 2279-543X | 1474 | Title of articles/authors |
| 24 | JAC: A Journal of Composition Theory | http://www.jctjournal.com/ | 0731-6755 | 1672 | Title of articles/authors/ affiliation |
| 25 | Journal for Advanced Research in Applied Sciences | http://iaetsdjaras.org/ | 2394-8442 | 947 | Title of articles/authors/ affiliation |
| 26 | Journal of Applied Science and Computations | http://j-asc.com/ | 1076-5131 | 3030 | Title of articles/authors/ affiliation |
| 27 | Journal of Computational Information Systems | http://jofcis.org/ | 1553-9105 (since 1 January 2021), 1548 - 7741 (till 31 December 2020) | 816 | Title of articles/authors/ affiliation |
| 28 | Journal of Engineering, Computing and Architecture | http://www.journaleca.com/ | 1934-7197 | 476 | Title of articles/authors/ affiliation |
| 29 | Journal of Huazhong University of Science and Technology | http://hxstxxjns.asia/ | 1671-4512 | 190 | Title of articles |
| 30 | Journal of Information and Computational Science | http://www.joics.org/ | 1548-7741 | 2056 | Title of articles/authors/ affiliation |
| 31 | Journal of Information and Computational Science | http://www.joics.net/ | 1548-7741 | 386 | Title of articles/authors |
| 32 | Journal of Interdisciplinary Cycle Research | http://www.jicrjournal.com/ | 0022-1945 | 1605 | Title of articles/authors/ affiliation |
| 33 | Journal of Interdisciplinary Cycle Research | http://positifreview.com/ | 0022-1945 | 391 | Title of articles/authors/ affiliation |
| 34 | Journal of Productivity Management | https://journalpm.com/ | 1868-8519 | 190 | Title of articles |
| 35 | Journal of Scientific Computing | http://jscglobal.org/ | 1524-2560 | 287 | Title of articles/authors/ affiliation |
| 36 | Journal of Shanghai Jiaotong University | https://shjtdxxb-e.cn/ | 1007-1172 | 158 | Title of articles |
| 37 | Journal of Southwest Jiaotong University | http://jsju.org/index.php/journal/index | 0258-2724 | 709 | Title of articles |
| 38 | Journal of University of Shanghai for Science and Technology | https://jusst.org/ | 1007-6735 | 172 | Title of articles |
| 39 | Journal of Xi'an Shiyou University, Natural Science Edition | https://www.xisdxjxsu.asia/ | 1673-064X | 203 | Title of articles/authors/ affiliation |
| 40 | Journal of Xi'an University of Architecture & Technology | http://www.xajzkjdx.cn/ | 1006-7930 | 1960 | Title of articles/authors/ affiliation |
| 41 | Journal of Xidian University | http://xadzkjdx.cn/ | 1001-2400 | 1391 | Title of articles |
| 42 | Mukt Shabd Journal | http://shabdbooks.com/ | 2347-3150 | 1804 | Title of articles/authors/ affiliation |
| 43 | Novyi Mir Research Journal | https://www.novyimir.net/ | 0130-7673 | 294 | Title of articles/authors/ affiliation |
| 44 | Paideuma Journal of Research | http://paideumajournal.com/ | 0090-5674 | 581 | Title of articles/authors/ affiliation |
| 45 | Parishodh Journal | http://www.parishodhpu.com/ | 2347-6648 | 626 | Title of articles/authors/ affiliation |
| 46 | Pensee | https://penseeresearch.com/ | 0031-4773 | 249 | Title of articles |
| 47 | Pramana Research Jour nal | https://www.pramanaresearch.org/ | 2249-2976 | 1534 | Title of articles/authors/ affiliation |
| 48 | Praxis Science and Technology Journal | http://praxisonline.org/ | 0369-8394 | 159 | Title of articles/authors/ affiliation |
| 49 | Proteus Journal | http://proteusresearch.org/ | 0889-6348 | 359 | Title of articles/authors/ affiliation |
| 50 | Science and Engineering Journal | https://saejournal.com/ | 0103-944X | 195 | Title of articles |
| 51 | Science, Technology and Development Multidisciplinary Journal | http://journalstd.com/ | 0950-0707 | 909 | Title of articles/authors/ affiliation |
| 52 | Science, Technology and Development Multidisciplinary Journal | http://dickensian.org/ | 0950-0707 | 346 | Title of articles/authors/ affiliation |
| 53 | Strad research | http://stradresearch.org/ | 0039-2049 | 417 | Title of articles/authors/ affiliation |
| 54 | Studia Rosenthaliana (Journal for The Study of Research) | http://www.jsrpublication.com/ | 1781-7838 | 346 | Title of articles |
| 55 | Suraj Punj Journal for Multidisciplinary Research | http://www.spjmr.com/ | 2394-2886 | 1685 | Title of articles/authors/ affiliation |
| 56 | Tierarztilich Praxis | http://tierarztliche.com/ | 0303-6286 | 198 | Title of articles |
| 57 | The International Journal of Analytical and Experimental Modal Analysis | http://www.ijaema.com/ | 0886-9367 | 3833 | Title of articles/authors/ affiliation |
| 58 | Universal Review | http://universalreview.org/* | 2277-2723 | 2491 | Title of articles |
| 59 | Vigyan Prakash Journal | http://lokvigyanparishad.com/ | 1549-523X | 407 | Title of articles/authors/ affiliation |
| 60 | Waffen- und Kostumkunde Journal | (https://www.druckhaus-hofmann.de/ | 0042-9945 | 338 | Title of articles/authors/ affiliation |
| 61 | Wutan Huatan Jisuan Jishu | http://www.wthtjsjs.cn/ | 1001-1749 | 338 | Title of articles/authors/ affiliation |
| 62 | Zeichen Journal | http://www.ezeichen.com/ | 0932-4747 | 289 | Title of articles/authors/ affiliation |

*- the website is not working as of 31 December 2020.

** - the website is not operational as of 31 December 2020.

## Appendix 2

# The list of clone journals

| N | Journal (clone) | URL (clone) | ISSN (clone) | Original journal | ISSN original |
|---|---|---|---|---|---|
| 1 | Adalya Journal | http://adalyajournal.com/ | 1301-2746 | Adalya | 1301-2746 |
| 2 | Aegaeum Journal | http://aegaeum.com/ | 0776-3808 | Aegaeum | 0776-3808 |
| 3 | Aegaum Journal | https://aegaum.com/ | 0776-3808 | Aegaeum | 0776-3808 |
| 4 | Alochana Chakra Journal | http://www.alochanachakra.in/* | 2231-3990 | Alochona Chakra | 2231-3990 |
| 5 | Aut Aut Research Journal | http://autrj.com/ | 0005-0601 | Aut Aut | 0005-0601 |
| 6 | Bulletin Monumental Journal | http://bulletinmonumental.com/ | 0007-473X | Bulletin Monumental | 0007-473X |
| 7 | Cikitusi Journal for Multidisciplinary Research | http://www.cikitusi.com/ | 0975-6876 | Cikitusi | 0975-6876 |
| 8 | Cithara Journal | http://citharajournal.com/ | 0009-7527 | Cithara | 0009-7527 |
| 9 | Compliance Engineering Journal | http://ijceng.com/ | 0898-3577 | Compliance Engineering | 0898-3577 |
| 10 | Degres Journal | https://degresjournal.com/ | 0376-8163 | Degrés | 0376-8163 |
| 11 | Dogo Rangsang Research Journal (D.R S. Research Journal) | http://www.drsrjournal.com/ | 2347-7180 | Dogo Rangsang Research Journal | 2347-7180 |
| 12 | Drewno Journal | https://drewnojournal.com/** | 1644-3985 | Drewno | 1644-3985 |
| 13 | G & O (Gedrag & Organisatie) | http://lemma-tijdschriften.com/ | 0921-5077 | Gedrag en Organisatie | 0921-5077, 1875-7235 |
| 14 | GIS Science Journal | http://www.gisscience.net/ | 1869-9391 | GIS.Science | 1869-9391 |
| 15 | Gorteria Journal | https://gorteria.com/ | 0017-2294 | Gorteria | 0017-2294 |
| 16 | High Technology Letters | http://www.gjstx-e.cn/ | 1006-6748 | High Technology Letters | 1006-6748 |
| 17 | Infokara Research | http://www.infokara.com/ | 1021-9056 | Infokara | 1021-9056 |
| 18 | International Journal of Information and Computing Science | http://ijics.com/ | 0972-1347 | International Journal of Information and Computing Science | 0972-1347 |
| 19 | International Journal of Innovative Research & Studies | http://ijirs.in/ | 2319-9725 | International Journal of Innovative Research and Studies | 2319-9725 |
| 20 | International Journal of Management, Technology and Engineering | http://www.ijamtes.org/ | 2249-7455 | International Journal of Advances in Management Technology and Engineering Sciences | 2249-7455 |
| 21 | International Journal of Research | http://ijrpublisher.com/ | 2236-6124 | International Journal of Research | 2236-6124 |
| 22 | International Journal of Scientific Research and Review | http://www.dynamicpublisher.org/ | 2279-543X | Les Cahiers de Chiasmi International (2279-543X), International Journal of Scientific Research and Reviews (2279-0543) | 2279-543X, 2279-0543 |
| 23 | International Journal of Scientific Research and Review | http://www.ijsrr.co.in/ | 2279-543X | Les Cahiers de Chiasmi International (2279-543X), International Journal of Scientific Research and Reviews (2279-0543) | 2279-543X, 2279-0543 |
| 24 | JAC: A Journal of Composition Theory | http://www.jctjournal.com/ | 0731-6755 | Journal of Composition Theory | 0731-6755 |
| 25 | Journal for Advanced Research in Applied Sciences | http://iaetsdjaras.org/ | 2394-8442 | Journal for Advanced Research in Applied Sciences | 2394-8442 |
| 26 | Journal of Applied Science and Computations | http://j-asc.com/ | 1076-5131 | Journal of Applied Science & Computations | 1076-5131 |
| 27 | Journal of Computational Information Systems | http://jofcis.org/ | 1553-9105 (since 1 January 2021), 1548 - 7741 (till 31 December 2020) | Journal of Computational Information Systems | 1553-9105 |
| 28 | Journal of Engineering, Computing and Architecture | http://www.journaleca.com/ | 1934-7197 | The Journal of Engineering, Computing and Architecture | 1934-7197 |
| 29 | Journal of Huazhong University of Science and Technology | http://hxstxxjns.asia/ | 1671-4512 | Huazhong Keji Daxue Xuebao/ Journal of Huazhong University of Science and Technology | 1671-4512 |
| 30 | Journal of Information and Computational Science | http://www.joics.org/ | 1548-7741 | Journal of Information and Computational Science | 1548-7741 |
| 31 | Journal of Information and Computational Science | http://www.joics.net/ | 1548-7741 | Journal of Information and Computational Science | 1548-7741 |
| 32 | Journal of Interdisciplinary Cycle Research | http://www.jicrjournal.com/ | 0022-1945 | Journal of Interdisciplinary Cycle Research (till 1993) | 0022-1945 |
| 33 | Journal of Interdisciplinary Cycle Research | http://positifreview.com/ | 0022-1945 | Journal of Interdisciplinary Cycle Research (till 1993) | 0022-1945 |
| 34 | Journal of Productivity Management | https://journalpm.com/ | 1868-8519 | Productivity Management | 1868-8519 |
| 35 | Journal of Scientific Computing | http://jscglobal.org/ | 1524-2560 | Scientific Computing | 1524-2560 |
| 36 | Journal of Shanghai Jiaotong University | https://shjtdxxb-e.cn/ | 1007-1172 | Journal of Shanghai Jiaotong University | 1007-1172 |
| 37 | Journal of Southwest Jiaotong University | http://jsju.org/index.php/journal/index | 0258-2724 | Xinan Jiaotong Daxue Xuebao/Journal of Southwest Jiaotong University | 0258-2724 |
| 38 | Journal of University of Shanghai for Science and Technology | https://jusst.org/ | 1007-6735 | Shanghai Ligong Daxue Xuebao/Journal of University of Shanghai for Science and Technology | 1007-6735 |
| 39 | Journal of Xi'an Shiyou University, Natural Science Edition | https://www.xisdxjxsu.asia/ | 1673-064X | Xi'an Shiyou Daxue Xuebao (Ziran Kexue Ban)/Journal of Xi'an Shiyou University, Natural Sciences Edition | 1673-064X |
| 40 | Journal of Xi'an University of Architecture & Technology | http://www.xajzkjdx.cn/ | 1006-7930 | Xi'an Jianzhu Ke-Ji Daxue xuebao/Journal of Xi'an University of Architecture & Technology | 1006-7930 |
| 41 | Journal of Xidian University | http://xadzkjdx.cn/ | 1001-2400 | Xi'an Dianzi Keji Daxue xuebao/Journal of Xidian University | 1001-2400 |
| 42 | Mukt Shabd Journal | http://shabdbooks.com/ | 2347-3150 | Mukt Shabd | 2347-3150 |
| 43 | Novyi Mir Research Journal | https://www.novyimir.net/ | 0130-7673 | Novyi Mir | 0130-7673 |
| 44 | Paideuma Journal of Research | http://paideumajournal.com/ | 0090-5674 | Paideuma | 0090-5674 |
| 45 | Parishodh Journal | http://www.parishodhpu.com/ | 2347-6648 | Parishodh | 2347-6648 |
| 46 | Pensee | https://penseeresearch.com/ | 0031-4773 | La Pensée | 0031-4773 |
| 47 | Pramana Research Journal | https://www.pramanaresearch.org/ | 2249-2976 | Parmana Research Journal | 2249-2976 |
| 48 | Praxis Science and Technology Journal | http://praxisonline.org/ | 0369-8394 | Praxis | 0369-8394 |
| 49 | Proteus Journal | http://proteusresearch.org/ | 0889-6348 | Proteus | 0889-6348 |
| 50 | Science and Engineering Journal | https://saejournal.com/ | 0103-944X | Ciência & Engenharia/Science and Engineering Journal | 0103-944X |
| 51 | Science, Technology and Development Multidisciplinary Journal | http://journalstd.com/ | 0950-0707 | Science, Technology& Development | 0950-0707 |
| 52 | Science, Technology and Development Multidisciplinary Journal | http://dickensian.org/ | 0950-0707 | Science, Technology& Development | 0950-0707 |
| 53 | Strad research | http://stradresearch.org/ | 0039-2049 | The Strad | 0039-2049 |
| 54 | Studia Rosenthaliana (Journal for The Study of Research) | http://www.jsrpublication.com/ | 1781-7838 | Studia Rosenthaliana | 1781-7838 |
| 55 | Suraj Punj Journal for Multidisciplinary Research | http://www.spjmr.com/ | 2394-2886 | Suraj Punj Journal | 2394-2886 |
| 56 | Tierarztilich Praxis | http://tierarztliche.com/ | 0303-6286 | Tierärztliche Praxis | 0303-6286 |
| 57 | The International Journal of Analytical and Experimental Modal Analysis | http://www.ijaema.com/ | 0886-9367 | The International Journal of Analytical and Experimental Modal Analysis | 0886-9367 |
| 58 | Universal Review | http://universalreview.org/* | 2277-2723 | Universal Review | 2277-2723 |
| 59 | Vigyan Prakash Journal | http://lokvigyanparishad.com/ | 1549-523X | Vijñana prakash (Vigyan prakash) | 1549-523X |
| 60 | Waffen- und Kostumkunde Journal | (https://www.druckhaus-hofmann.de/ | 0042-9945 | Waffen- und Kostümkunde | 0042-9945 |
| 61 | Wutan Huatan Jisuan Jishu | http://www.wthtjsjs.cn/ | 1001-1749 | Wu-tan hua-tan jisuan jishu | 1001-1749 |
| 62 | Zeichen Journal | http://www.ezeichen.com/ | 0932-4747 | Zeichen | 0932-4747 |

*- the website is not working as of 31 December 2020.
** - the website is not operational as of 31 December 2020.